\documentstyle[sprocl]{article}

\def\la{\langle }
\def\ra{ \rangle }
 \def\gmf{\gamma _{5}}
\def\beq{\begin{equation}}
\def\eeq{\end{equation}}
\def\bea{\begin{eqnarray}}
\def\eea{\end{eqnarray}}
\begin{document}

\title{PRIMORDIAL GALACTIC MAGNETIC FIELDS FROM THE QCD PHASE
  TRANSITION} 
\author{ Michael Forbes and Ariel Zhitnitsky}
\address{Department of Physics and Astronomy, University of British Columbia\\
  Vancouver, British Columbia, Canada V6T 1Z1.}  
\maketitle
\abstracts{In this letter, we propose a mechanism to generate
  large-scale magnetic fields with correlation lengths of $100$ kpc.
  Domain walls with QCD scale internal structure form and coalesce
  obtaining Hubble scale correlations and align nucleon spins.  Due to
  strong CP violation, nucleons in these walls have anomalous electric
  and magnetic dipole moments and thus the walls are ferromagnetic.
  This induces electromagnetic fields with Hubble size correlations.
  The same CP violation also induces a maximal helicity (Chern-Simons)
  correlated through the Hubble volume which supports an ``inverse
  cascade'' allowing the initial correlations to grow to $100$ kpc
  today.  We estimate the generated electromagnetic fields in terms of
  the QCD parameters and discuss the effects of the resulting fields.}
\section{Introduction}
The source of cosmic magnetic fields with large scale correlations has
remained somewhat of a mystery \cite{observations}.  There are two
possible origins for these fields: primordial sources and galactic
sources.  Primordial fields are produced in the earlier universe, then
evolve, and are thought to provide seeds which gravitational dynamos
later amplify.  Galactic sources would produce the fields as well as
amplify them.  Many mechanisms have been
proposed \cite{magreview,Cornwall:1997,Son:1999,FC:1998}, however, most
fail to convincingly generate fields with large enough correlation
lengths to match the observed microgauss fields with $\sim 100$ kpc
correlations.  We present here a mechanism which, although probably
requiring a dynamo to produce microgauss fields, generates fields with
hundred kiloparsec correlations.  We present this mechanism as an
application of our recent understanding of QCD domain walls, which
will be described in detail elsewhere \cite{FZ:2000}.

\begin{enumerate}
  \setlength{\parskip}{-3pt}
\item Sometime near the QCD phase transition, $T_{\mathrm{QCD}}\approx
  1$ GeV, QCD domain walls form.
\item These domain walls rapidly coalesce until there remains, on
  average, one domain wall per Hubble volume with Hubble scale
  correlations.
\item Baryons interact with the domain walls and align their spins
  along the domain walls.
\item The magnetic and electric dipole moments of the baryons induce
  helical magnetic fields correlated with the domain wall.
\item The domain walls decay, leaving a magnetic field.
\item As the universe expands, an ``inverse cascade'' mechanism
  transfers energy from small to large scale modes, effectively
  increasing the resulting correlation length of the observed large
  scale fields.
\end{enumerate}
We shall start by discussing the ``inverse cascade'' mechanism because
it seems to be the most efficient mechanism for increasing the
correlation length of magnetic turbulence.  After presenting some
estimates to show that this mechanism can indeed generate fields of
the observed scales, we shall discuss the domain wall mechanism for
generating the initial fields and some relevant astrophysical
phenomena associated with this mechanism.

\section{Evolution of Magnetic Fields}
As suggested by Cornwall \cite{Cornwall:1997}, discussed by Son
\cite{Son:1999} and confirmed by Field and Carroll \cite{FC:1998},
energy in magnetic fields can undergo an apparent ``inverse cascade''
and be transfered from high frequency modes to low frequency modes,
thus increasing the overall correlation length of the field faster
than the na\"\i{}ve scaling by the universe's scale parameter $R(T)$.
There are two important conditions: turbulence must be supported as
indicated by a large Reynolds number ${\rm Re}$, and magnetic helicity
(Abelian Chern-Simons number) $H=\int{\vec{\bf A}\cdot\vec{\bf B}}{\rm
  d}^3{x}$ is approximately conserved.  The importance of helicity was
originally demonstrated by Pouquet and collaborators \cite{Pouquet}.
The mechanism is thus: the small scale modes dissipate, but the
conservation of helicity requires that the helicity be transfered to
larger scale modes. Some energy is transfered along with the helicity
and hence energy is transported from the small to large scale modes.
This is the ``inverse cascade''.\cite{Cornwall:1997,Son:1999,FC:1998}

In the early universe, ${\rm Re}$ is very large and supports
turbulence.  This drops to $\mathrm{Re}\approx 1$ at the $e^+e^-$
annihilation epoch,\cite{Son:1999} $T_0\approx 100$ eV.  After this
point (and throughout the matter dominated phase) we assume that the
fields are ``frozen in'' and that the correlation length expands as
$R$ while the field strength decays as $R^{-2}$.  Note that the
``inverse cascade'' is only supported during the radiation dominated
phase of the universe.

Under the assumption that the field is maximally helical, these
conditions imply the following relationships \cite{Son:1999,FC:1998}
between the initial field $B_{\mathrm{rms}}(T_i)$ with initial
correlation $l(T_i)$ and present fields today
($T_{\mathrm{now}}\approx2\times10^{-4}$ eV)
$B_{\mathrm{rms}}(T_{\mathrm{now}})$ with correlation
$l(T_{\mathrm{now}})$:
\begin{eqnarray}
  B_{\mathrm{rms}}(T_{\mathrm{now}})&=&
  \left(\frac{T_0}{T_{\mathrm{now}}}\right)^{-2}
  \left(\frac{T_i}{T_0}\right)^{-7/3}B_{\mathrm{rms}}(T_i)
  \label{eq:brms1}\\
  l(T_{\mathrm{now}})&=&\left(\frac{T_0}{T_{\mathrm{now}}}\right)
  \left(\frac{T_i}{T_0}\right)^{5/3}l(T_i).
  \label{eq:corr1}
\end{eqnarray}

As pointed out by Son \cite{Son:1999}, the only way to generate
turbulence is either by a phase transition $T_i$ or by gravitational
instabilities.  We consider the former source.  As we shall show, our
mechanism generates Hubble size correlations $l_i$ at a phase
transition $T_i$.  In the radiation dominated epoch, the Hubble size
scales as $T_i^{-2}$.  Combining this with (\ref{eq:corr1}), we see
that $l_{\mathrm{now}}\propto T_i^{-1/3}$; thus, the earlier the phase
transition, the smaller the possible correlations.

The last phase transition is the QCD transition,
$T_i=T_{\mathrm{QCD}}\approx 0.2$ GeV with Hubble size
$l(T_{\mathrm{QCD}})\approx 30$ km.  We calculate (\ref{eq:EB}) the
initial magnetic field strength to be $B_{\mathrm{rms}}(T_i)\approx
e\Lambda_{QCD}^2/(\xi\Lambda_{\rm QCD})\approx (10^{17}{\rm
  G})/(\xi\Lambda_{\rm QCD})$ where $\xi$ is a correlation length that
depends on the dynamics of the system as discussed below and
$\Lambda_{\rm QCD}\approx 0.2$ GeV.  With these estimates, we see that
we can achieve
\begin{equation}
  \label{*}
  B_{\mathrm{rms}}\sim
  \frac{10^{-9}{\rm G}}{\xi\Lambda_{\rm QCD}} ,~~~~ l\sim 100~{\rm kpc}
\end{equation}
today.  One might consider the electroweak transition which might
produce $100$ pc correlations today, but this presupposes a mechanism
for generating fields with Hubble scale correlations.  Such a
mechanism does not appear to be possible in the Standard Model.
Instead, the fields produced are correlated at the scale $T_i^{-1}$
which can produce only $\sim 1$ km correlations today.

These are crude estimates, and galactic dynamos likely amplify these
fields.  The important point is that we can generate easily the $100$
kpc correlations observed today {\em provided} that the fields were
initially of Hubble size correlation.  Unless another mechanism for
amplifying the correlations of magnetic fields is discovered, we
suggest that, in order to obtain microgauss fields with $100$ kpc
correlation lengths, helical fields must be generated with Hubble
scale correlations near or slightly after the QCD phase transition
$T_{\mathrm{QCD}}$.  The same conclusion regarding the relevance of
the QCD scale for this problem was also reached by Son, Field and
Carroll \cite{Son:1999,FC:1998}.  The rest of this work presents a
mechanism that can provide the desired Hubble size fields, justifying
the estimate (\ref{*}).  We shall explain the mechanism and give
simple estimates here, but present details of the calculations
elsewhere \cite{FZ:2000}.

\section{Domain Walls}
The key players in our mechanism are domain walls formed at the QCD
phase transition that possess an internal structure with QCD scale.
We shall present a full exposition about these types of walls in
another paper \cite{FZ:2000} but, to be specific, here we shall
discuss the so-called axion-$\eta'$ ($a_{\eta'}$) domain wall
\cite{FZ:2000}.
 
We start with a similar effective Lagrangian to that used by Huang and
Sikivie except that we included the effects of the $\eta'$ singlet
field which they neglected.  The Lagrangian density is
\begin{equation}
  \label{eq:leff}
  {\cal L}_{{\rm eff}}=
   \frac{f_a^2}{2}\left|\partial_\mu e^{i {a}}\right|^2+
   \frac{f_{\pi}^2}{4}{\rm Tr}\left|\partial_\mu{\bf U}\right|^2-
   V({\bf U}, {a})
\end{equation}
where $a$ is the dimensionless axion field and the matrix ${\bf
  U}=\exp(i {\eta}'+i {\pi}^f{\bf \lambda}^f)$ contains the pion and
$\eta'$ fields (to simplify the calculations, we consider only the
$SU(2)$ flavor group).  The $\eta'$ field is not light, but as we
shall see, is the dominant player in aligning the magnetic fields so
we include it.  The potential $V$ is given by
\begin{equation}
  \label{potential}
  V =\frac{1}{2} {\rm Tr}\left({\bf MU}e^{ i {a }}+~{\rm h.c.}\right)-
  E\cos\left(\frac{i\ln[\det({\bf U})]}{
      N_c}\right)
\end{equation}
which was first introduced by Halpern and Zhitnitsky.\cite{HZ} It
should be realized that $i\ln[\det({\bf U})]\equiv i\ln(\det({\bf
  U}))+2\pi n$ is a multivalued function and we must choose the
minimum valued branch.  Details about this potential are discussed in
the original paper \cite{HZ} but several points will be made here.
All dimensionful parameters are expressed in terms of the QCD chiral
and gluon vacuum condensates and are well known numerically: ${\bf M}
= -{\rm diag} (m_{q}^{i} |\langle \bar{q}^{i}q^{i} \rangle| )$ and $ E
= \langle b \alpha_s /(32 \pi) G^2 \rangle$.

The result is that two different types of axion domain walls
form.\cite{FZ:2000} One is almost identical to the one discussed by
Huang and Sikivie \cite{HS:1985} with small corrections due to the
$\eta'$ field.  We shall call this the axion/pion ($a_{\pi}$) domain
wall.  The second type, which we shall call the axion/eta'
($a_{\eta'}$) domain wall is a new solution characterized by a
transition in both the axion and $\eta'$ fields (see our other paper
\cite{FZ:2000} for a complete description of this wall).  The boundary
conditions (vacuum states) for this wall are
${a}(-\infty)={\eta}'(-\infty)=0$ and
${a}(\infty)={\eta}'(\infty)=\pm\pi$ with $\pi^0=0$ at
both boundaries.  

The main difference between the structures of the two walls is that,
whereas the $a_{\pi}$ domain wall has structure only on the huge scale
of $m_a^{-1}$, the $\eta'$ transition in the $a_{\eta'}$ has both
scales, the axion scale, $m_a^{-1}$, as well as $\Lambda_{\rm
  QCD}^{-1}$ scale.  Therefore, the $a_{\eta'}$ domain wall has a
``sandwich'' structure.  The reason is that, in the presence of the
non-zero axion ($\theta$) field, the pion becomes effectively massless
due to its Goldstone nature.  The $\eta'$ is not so sensitive to
$\theta$ parameter and so its mass never becomes zero.  It is crucial
that the walls have a structure of scale $\Lambda_{\rm QCD}^{-1}$:
thus there is no way for the $a_{\pi}$ wall to trap nucleons because
of the huge differece in scales but the $a_{\eta'}$ wall has exactly
this structure and can therefore efficiently align the nucleons. The
QCD domain walls (which were also discussed in \cite{FZ:2000}) have
the same property as $a_{\eta'}$ walls.  Namely, they have the
structure of scale $\Lambda_{\rm QCD}^{-1}$ and they can play the same
role as $a_{\eta'}$ walls.  In what follows, for more concreteness, we
use $a_{\eta'}$ walls.
 
The model we propose is this: Immediately after the phase transition,
the universe is filled with domain walls on the scale of $T_{\rm QCD
  }^{-1}$.  As the temperature drops, these domain walls coalesce,
resulting in an average of one large domain wall per Hubble volume
with Hubble scale correlations \cite{CHS:1999,strings}.  It is these
Hubble scale $a_{\eta'}$ domain walls which align the dipole moments
of the nucleons producing the seed fields.

The following steps are crucial for this phenomenon:
\begin{enumerate}
  \setlength{\parskip}{-3pt}
\item The coalescing of QCD domain wall gives the fields
  $\pi,\dots,\eta'$ Hubble scale correlations.
\item These fields interact with the nucleons producing Hubble scale
  correlations of nucleon spins residing in the vicinity of the domain
  wall.  (The spins align perpendicular to the wall surface.)
\item Finally, the nucleons, which carry electric and magnetic moments
  (due to strong CP violation), induce Hubble scale correlated
  magnetic and electric fields.
\item These magnetic and electric fields eventually induce a nonzero
  helicity which has the same correlation.  This helicity enables the
  inverse cascade.
\end{enumerate}
\section{  Domain Wall Properties}
We present here a method for simplifying the calculations of the bulk
properties of domain walls.  This method makes the approximation that
the domain wall is flat and that translational and rotational
symmetries are preserved in the plane of the wall which we take to be
the $x$--$y$ plane.  These approximations are valid in the case of
domain walls whose curvature is large in comparison to the length
scale of the pertinent physics.

Once this approximation is made, we can reformulate the problem in
$1+1$ dimensions ($z$ and $t$) and calculate the density of the
desired bulk properties along the domain wall.  To regain the full
four-dimensional bulk properties, we must estimate the density of the
particles in the $x$--$y$ plane to obtain the appropriate density and
degeneracy factors for the bulk density.  Thus, the final results are
not independent of physics in the $x$--$y$ plane, but rather, these
effects are accounted for only through the degeneracy factors.

\subsection{Alignment of Spins in the Domain Wall Background}
We proceed to demonstrate this technique by calculating the alignment
of fermionic spins along the wall.  To estimate the strengths of the
fields involved, we consider only the $\eta'$ transition because it
has a similar structure in both the $a_{\eta'}$ domain wall and the
QCD domain walls.  We take the following simple interaction between
the $\eta'$ field and the nucleons:
\begin{equation}
  \label{4d}  \bar{\Psi}\left[i{\not{\!\partial}}-m_Ne^{i {\eta}'
      (z)\gamma_5}\right]\Psi.
\end{equation}
For our approximations, we assume that fluctuations in the nucleon
fields do not affect the domain walls and, thus, treat the domain
walls as a background field\footnote{A full account would take into
  account the effects of this back-reaction.  We expect that they
  would affect the potential (\ref{potential}) by altering the form of
  the last term $E\cos()$ and possibly adding higher order
  corrections, but that they would not alter the nature of the domain
  walls.  Quantitatively this would alter the numerical results, but
  would not change the qualitative picture presented here.}.  The strategy is to break (\ref{4d})
into two $1+1$ dimensional components by setting
$\partial_x=\partial_y=0$ and then by manipulating the system of
equations that result to obtain an equivalent two-dimensional system.

First, we introduce the following chiral components of the Dirac
spinors
\begin{equation}
  \label{ad4}
  \Psi_{+} \equiv \left( \begin{array}{c}
      \chi_{1} \\ \chi_{2} \end{array} \right),~~
  \Psi_{-} \equiv \left( \begin{array}{c}
      \eta_{1} \\ \eta_{2} \end{array} \right),~~
  \Psi 
  = \frac{1}{\sqrt{2}}\left(
    \begin{array}{c}
      \Psi_{+}+\Psi_{-}\\
      \Psi_{+}-\Psi_{-}  
    \end{array}
  \right)
\end{equation}
Secondly, we assume that our system is effectively two-dimensional (an
infinite domain wall lying in the $x$--$y$ plane) and hence neglect
the conserved $x$ and $y$ momenta.  We find that the Dirac equations
which follow from (\ref{ad4}) are equivalent to the coupled system
\begin{eqnarray}
  \label{ad5}
  \left[ i \partial_0 + i \sigma_3 \partial_3  
  \right] \Psi_{+} = m_{N} e^{-i \eta'} \Psi_{-} \nonumber \\
  \left[ i \partial_0 - i \sigma_3 \partial_3  
  \right] \Psi_{-} = m_{N} e^{+i \eta'} \Psi_{+}
\end{eqnarray}
in $\Psi_{+}$ and $\Psi_{-}$.  At this stage it proves convenient to
rearrange these equations by introducing new ``two-dimensional'' Dirac
spinors $\Psi^{(1)}$ and $\Psi^{(2)}$
  \begin{eqnarray}
  \label{ad6}
  \Psi^{(1)} =   \left( 
    \begin{array}{c}
      \chi_{1} \\ \eta_{1}
    \end{array}
  \right), \; 
  \Psi^{(2)} = \left(
    \begin{array}{c}
      \eta_{2} \\ \chi_{2}
    \end{array}
  \right)
\end{eqnarray}
Using the definitions (\ref{ad6}), we put system (\ref{ad5}) into the
form of two two-dimensional (2D) Dirac equations
\begin{eqnarray}
  \label{ad7}
  \left[ i \hat{\gamma}_{\mu} \partial_{\mu}   
    - m_{N} e^{-i \eta' \hat{\gmf} }\right] \Psi^{(1)} = 0 \nonumber \\
  \left[ i \hat{\gamma}_{\mu} \partial_{\mu}  
    - m_{N} e^{+i \eta' \hat{\gmf} }\right] \Psi^{(2)} = 0   
\end{eqnarray}
where we have introduced the 2D Dirac matrices
\begin{equation}
  \label{ad8}
  \hat{\gamma}_0 = \sigma_1,
  \;\; \hat{\gamma}_1 = -i \sigma_{2},
  \;\; \hat{\gamma}_5 = \hat{\gamma}_0 \hat{\gamma}_1 = \sigma_3,
  \;\;
  \hat{\gamma}_{\mu}\hat{\gamma}_{\nu}=g_{\mu\nu}+\epsilon_{\mu\nu}
  \hat{\gamma}_5.
\end{equation}
Equations (\ref{ad7}) are reproduced from the following effective 2D 
Lagrangian
\begin{eqnarray}
  \label{ad9}
  {\cal L}_{\rm 2D} = \bar{\Psi}^{(1)} \left[ i \hat{\gamma}_{\mu} 
    \partial_{\mu}   - 
    m_{N} e^{i \eta' \hat{\gmf} }
  \right] \Psi^{(1)} 
  +  \bar{\Psi}^{(2)} \left[ i \hat{\gamma}_{\mu} 
    \partial_{\mu}  - 
    m_{N} e^{-i \eta' \hat{\gmf} }
  \right] \Psi^{(2)}, 
\end{eqnarray} 
where $\eta'=\eta'(z)$ is the background classical field with boundary
conditions $\eta'(-\infty)=0$, $\eta'(\infty)=\pi$.  This 2D
Lagrangian describes two species of 2D Dirac fermions of oposite
chiral charge intracting with the external $\eta'$ field.

Now we are ready to demonstrate that the domain walls align the spins
of the fermions.  The relevant operator (which becomes the spin
operator for the nonrelativistic system) is
\begin{eqnarray}
  \label{ad10}
  {\Psi^{\dagger}\vec{\Sigma}\Psi  =\bar{\Psi}    \vec{\gamma}
    \gamma_5}  \Psi 
  = \Psi_{+}^{\dagger} \vec{\sigma } \Psi_{+}
  +\Psi_{-}^{\dagger} \vec{\sigma}  \Psi_{-} ,~~ 
  \vec{\Sigma}\equiv \left( \begin{array}{c}
      \vec{\sigma} ~~ 0 \\ 0~~ \vec{\sigma} \end{array} \right),
\end{eqnarray}
and our goal is to demonstrate that the mean value $\la \Sigma_z\ra$
of this operator is generally non-zero in the domain wall.  Thus, the
nucleon spins are aligned in the $z$ direction and have a correlation
length similar to the domain wall.
 
The easiest way to demonstrate this phenomenon in our model (which is
effectively 2D) is to use the Goldstone-Wilczek adiabatic
approximation \cite{JR,GW} together with a bosonization trick
\begin{equation}
  \label{ad11}
  \begin{array}{cc}
  \bar{\Psi}^{(i)} i \hat{\gamma}_{\mu} \partial_{\mu} \Psi^{(i)}
  \rightarrow 
  \frac{1}{2} \partial_{\mu} \phi_{i} \partial_{\mu} \phi_{i}
  &
  \qquad\bar{\Psi}^{(i)} i \hat{\gamma}_5  \Psi^{(i)} 
  \rightarrow -{\mu}  \sin  \left( 2 
    \sqrt{\pi} \phi_i  \right)
  \\
  ~~~~~\bar{\Psi}^{(i)} \hat{\gamma}_{\mu} \Psi^{(i)}
  \rightarrow \frac{1}{\sqrt{\pi}} \varepsilon_{ \mu \nu} \partial_{\nu}
  \phi_{i}
  &
  \qquad~~~~\bar{\Psi}^{(i)} \Psi^{(i)} 
  \rightarrow  -{\mu}  \cos \left( 2 
    \sqrt{\pi} \phi_i  \right)   
\end{array}
\end{equation}
($\mu\sim m_N$ is a scale parameter).  The Lagrangian (\ref{ad9})
after bosonization is
\begin{eqnarray}
  \label{ad12}
  {\cal L}_{\rm 2D} &=& \frac{1}{2} \left[ 
    (\partial_{\mu} \phi_1 )^2 + (\partial_{\mu} \phi_2)^2 
  \right]- U(\phi_{1},\phi_{2} ) \nonumber \\
  U(\phi_{1},\phi_{2} ) &=& - \mu m_{N} \left[ 
    \cos (2 \sqrt{\pi} \phi_1 - \eta') + 
    \cos (2 \sqrt{\pi} \phi_2 +  \eta') \right]  
\end{eqnarray}
The adiabatic approximation \cite{GW} is to neglect the kinetic terms
in the analysis of the dynamics of $\phi_1$ and $\phi_2$ fields in
Equation (\ref{ad12}).  In this case, the mean values
$\la\phi_1(z)\ra$ and $\la \phi_2(z)\ra$ will follow the background
field $\eta'(z)$, and can be found by minimizing the potential $U$ in
Equation (\ref{ad12}).
\begin{eqnarray}
  \label{ad15}
  \la \phi_{1}(z)\ra = \frac{\eta' (z)}{2 \sqrt{\pi}}, ~~~ 
  \la\phi_{2}(z)\ra =- \frac{\eta'(z) }{2 \sqrt{\pi}}  
\end{eqnarray}
To calculate the induced spin $\Psi^{\dagger}\vec{\Sigma}\Psi$ in our
theory we should present the spin operator in terms of 2D fields
$\phi_{1}$ and $\phi_{2}$, and replace these fields by their mean
values (\ref{ad15}) in the domain wall background:
\begin{eqnarray}
  \label{ad16}
  \Psi^{\dagger} {\Sigma_z}\Psi  = \bar{\Psi}^{(1)} \hat{\gamma}_{0}
  \Psi^{(1)}-\bar{\Psi}^{(2)} \hat{\gamma}_{0}
  \Psi^{(2)}=\frac{1}{ \sqrt{\pi}}\left(\partial_z\phi_1(z)-\partial_z\phi_2(z)\right),
\end{eqnarray}
where, in the last step, we used the bosonic representation for the 2D
$\Psi^{(i)}$ fields\footnote{The 2D problem under discussion is quite
  familiar to physics community: namely, the calculation of the
  induced fermion charge in a solitonic background.}. The last step is
to replace these fields by their mean-values (\ref{ad15})
\begin{eqnarray}
  \label{ad17}
  \la\Psi^{\dagger} {\Sigma_z}\Psi \ra = N\times
  \frac{1}{  \pi} \frac{\partial\eta'(z)}{\partial z},
\end{eqnarray}
where $N$ is the appropriate normalization and degeneracy factor for
the $\Psi$ field which has canonical dimension $3/2$ in four
dimensions while the 2D $\Psi^{(i)}$ fields have canonical dimension
$1/2$.
  
\subsection{Fermion  Degeneracy in the Domain wall Background}
We have assumed that locally the domain walls have only a spatial $z$
dependence.  This implies  that there is still a $2$-dimensional
translational and rotational symmetry in the $x$--$y$ plane.  These
translational degrees of freedom imply that momentum in the plane is
conserved and hence we can treat the neglected degrees of freedom for
the fermions as free degrees.  The degeneracy in a region of area
$S$ will simply be a sum over these degrees with a discrete factor
$g=4=2\times2$ for spin and isospin degeneracy
\begin{eqnarray}
  \label{ad18}
  N=g\int\frac{dxdydp_xdp_y}{(2\pi)^2}=\frac{gp_F^2}{4\pi}S\simeq \frac{\Lambda_{QCD}^2}{\pi}S
\end{eqnarray}
where we estimated the Fermi energy $p_F\simeq \Lambda_{QCD}\simeq
150$ MeV.  As expected, the degeneracy is proprtional to the area of
the domain wall $S$.  Now it is clear that an appropriate
normalization for the two dimensional $\Psi^{(i)}$ fields can be
achieved by adding a factor $1/\sqrt{S}$ in the definition
(\ref{ad6}). In this case these 2D fields have correct canonical
dimension $1/2$. Now we are ready to estimate the original
four-dimensional expectation value (\ref{ad17}):
\begin{eqnarray}
\label{ad20}
\la \Psi^{\dagger} {\Sigma_z}\Psi\ra_{{\rm 4D}} =   \frac{1}{S}\times N\times
\frac{1}{  \pi} \frac{\partial\eta'(z)}{\partial z}
\simeq \frac{\Lambda_{QCD}^2}{\pi^2}\frac{\partial\eta'(z)}{\partial z}
\sim {\frac{\Lambda_{QCD}^2m_{\eta'}}{\pi^2}} ,
\end{eqnarray} 
which has correct dimension 3.  Using the same technique one can
estimate other matrix elements in the domain wall background which
have non-zero magnitude and thus demonstrate that they have a large
correlation $L$ on the of the size of the domain wall in the $x$--$y$
direction.  In particular, the result for the mean value
$\langle\bar{\Psi}\gamma_5\sigma_{xy}\Psi\rangle$ is:
\begin{equation}
  \label{em3}
  \la\bar{\Psi} \sigma_{xy}\gamma_{5}\Psi\ra_{\rm 4D} \sim
  \mu{\frac{\Lambda_{QCD}^2}{\pi}},
\end{equation}
where the factor $\sim \Lambda_{QCD}^2/\pi$ has the same origin as in
Equation (\ref{ad20}) and is related to the degeneracy of the system
(\ref{ad18}), while the factor $\mu\sim m_N$ is a dimensional
parameter originating from the bosonic representation (\ref{ad12}) of
the effective two-dimensional theory.

\section{Magnetic Field Generation Mechanism}
Here we estimate the strengths of the induced fields in terms of the
QCD parameters.  We consider two types of interactions.  First, the
nucleon spins align with the domain wall.  We assume that the
fluctuations in the nucleon field $\Psi$ are rapid and that these
effects cancel, leaving the classical domain wall background
unaltered.  Thus, we are able to estimate many mean values correlated
on a large scale on the domain walls such as
$\langle\bar{\Psi}\gamma_5\sigma_{xy}\Psi\rangle$ (\ref{em3}) and
$\langle\bar{\Psi}\gamma_z\gamma_5\Psi\rangle$ (\ref{ad20}) through
these interactions as described above.  These mean values are only
nonzero within a distance $\Lambda_{\rm QCD}^{-1}$ of the domain wall
and are correlated on the same Hubble scale as the domain wall.

From now on we treat the expectation value (\ref{em3}) as a background
classical field correlated on the Hubble scale.  Once these sources
are known, one could calculate the generated electromagnetic field by
solving Maxwell's equations with the interaction
\begin{equation}
  \label{em4}
  {\cal L}_{\rm int}={\frac{1}{2}}(d_{\Psi}\bar{\Psi}
  \sigma_{\mu\nu}\gamma_5\Psi +\mu_{\Psi}\bar{\Psi}i \sigma_{\mu\nu}
  \Psi) F_{\mu\nu} + \bar{\Psi}(i{\rm D})^2\Psi
\end{equation}
where $d_{\Psi}$ ($\mu_{\Psi}$) is effective electric (magnetic)
dipole moments of the field $\Psi$.  Due to the CP violation (nonzero
$\theta$) along the axion domain wall, the anomalous nucleon dipole
moment in (\ref{em4}) $d_{\Psi}\sim \mu_{\Psi}\sim \frac{e}{m_N}$ is
also nonzero \cite{CDVW:1979}.  This is an important point: if no
anomalous moments were induced, then only charged particles could
generate the magnetic field: the walls would be diamagnetic not
ferromagnetic as argued by Voloshin \cite{Voloshin2} and Landau levels
would exactly cancel the field generated by the dipoles.

Solving the complete set of Maxwell's equations, however, is extremely
difficult.  Instead, we use simple dimensional arguments.  For a small
planar region of area $\xi^2$ filled with aligned dipoles with
constant density, we know that the net magnetic field is proportional
to $\xi^{-1}$ since the dipole fields tend to cancel, thus for a flat
section of our domain wall, the field would be suppressed by a factor
of $(\xi\Lambda_{\rm QCD})^{-1}$.  For a perfectly flat, infinite
domain wall ($\xi\rightarrow \infty$), there would be no net field as
pointed out \cite{Voloshin2}.  However, our domain walls are far from
flat.  Indeed, they have many wiggles and high frequency modes, thus,
the size of the flat regions where the fields are suppressed is
governed by a correlation $\xi$ which describes the curvature of the
wall.  Thus, the average electric and magnetic fields produced by the
domain wall are of the order
\begin{equation}
  \label{em5}
  \langle F_{\mu\nu}\rangle \simeq
  \frac{1}{\xi^*\Lambda_{\rm QCD}}\left[d_{\Psi}\langle \bar{\Psi}
    \sigma_{\mu\nu}\gamma_{5}\Psi\rangle +\mu_{\Psi}\langle \bar{\Psi}i
    \sigma_{\mu\nu} \Psi\rangle\right]
\end{equation}
where $\xi^*$ is an effective correlation length related to the size
of the dominant high frequency modes.

To estimate what effective scale $\xi^*$ has, however, requires an
understanding of the dynamics of the domain walls.  Initially, the
domain walls are correlated with a scale of $\Lambda_{\rm QCD}^{-1}$.
As the temperature cools, the walls smooth out and the lower bound
$\xi(t)$ for the scale of the walls correlations increases from
$\xi(0)\simeq\Lambda_{\rm QCD}^{-1}$.  This increase is a dynamical
feature, however, and is thus slow.  In addition, the walls coalesce
and become correlated on the Hubble scale generating large scale
correlations.  Thus the wall has correlations from $\xi(t)$ up to the
upper limit set by the Hubble scale.  Thus, the effective
$\xi^*\ll$~Hubble~size at the time that the fields are aligned and so
the suppression is not nearly as great as implied by Voloshin
\cite{Voloshin2}.  Note that, even though the effects are confined to
the region close to the wall, the domain walls are moving and twisted
so that the effects occur throughout the entire Hubble volume.

The picture is thus that fields of strength
\begin{equation}
  \label{eq:EB}
  \langle E_z\rangle\simeq\langle B_z\rangle\sim 
  \frac{1}{\xi^*\Lambda_{\rm QCD}}\frac{e}{m_N}\frac{m_N\Lambda_{\rm QCD}^2}{\pi}
  \sim
  \frac{e\Lambda_{\rm QCD}}{\xi^*\pi}
\end{equation}
are generated with short correlations $\xi^*$, but then domains are
correlated on a large scale by the Hubble scale modes of the
coalescing domain walls.  Thus, strong turbulence is generated with
correlations that run from $\Lambda_{\rm QCD}$ up to the Hubble scale.

Finally, we note that this turbulence should be highly helical.  This
helicity arises from the fact that both electric and magnetic fields
are correlated together along the entire domain wall, $\langle\vec{\bf
  E}\rangle\sim\langle \vec{\bf A}\rangle/\tau$ where $\langle
\vec{\bf A}\rangle$ is the vector potential and $\tau$ is a relevant
timescale for the electrical field to be screened (we expect
$\tau\sim\Lambda_{\rm QCD}^{-1}$ as we discuss below).  The magnetic
helicity density is thus:
\begin{equation}
  \label{helicity}
  h\sim{\vec{\bf A}\cdot\vec{\bf B}} \sim\tau\langle E_z \rangle\langle
  B_z \rangle \sim \tau\frac{e^2}{\pi^2}\frac{\Lambda_{QCD}^2}{{\xi^*}^2}.
\end{equation}
Note carefully what happens here: The total helicity was zero in the
quark-gluon-plasma phase and remains zero in the whole universe, but
the helicity is separated so that in one Hubble volume, the helicity
has the same sign.  The reason for this is that, as the domain walls
coalesce, initial perturbations cause either a soliton or an
antisoliton to dominate and fill the Hubble volume.  In the
neighboring space, there will be other solitons and antisolitons so
that there is an equal number of both, but they are separated and this
spatial separation prevents them from annihilating.  This is similar
to how a particle and anti-particle may be created and then separated
so they do not annihilate.  In any case, the helicity is a
pseudoscalar and thus maintains a constant sign everywhere along the
domain wall: thus, the entire Hubble volume is filled with helicity of
the same sign.  This is the origin of the Hubble scale correlations in
the helicity and in $B^2$.  The correlation parameter $\xi$ which
affects the magnitude of the fields plays no role in disturbing this
correlation.

As we mentioned, eventually, the electric field will be screened.  The
timescale for this is set by the plasma frequency for the electrons
(protons will screen much more slowly) $\omega_p$ which turns out to
be numerically close to $\Lambda_{\rm QCD}$ near the QCD phase
transition.  The nucleons, however, also align on a similar timescale
$\Lambda_{QCD}^{-1}$, and the helicity is generated on this scale too,
so the electric screening will not qualitatively affect the mechanism.
Finally, we note that the turbulence requires a seed which remains in
a local region for a timescale set by the conductivity
\cite{Dimopoulos:1997nq} $\sigma\sim cT/e^2\sim\Lambda_{\rm QCD}$
where for $T=100$ MeV, $c\approx 0.07$ and is smaller for higher $T$.
Thus, even if the domain walls move at close to the speed of light
(due to vibrations), there is still enough time to generate
turbulence.

For this mechanism to work and not violate current observations, it
seems that the domain walls must eventually decay.  Several mechanisms
have been discussed for the decay of axion domain walls
\cite{CHS:1999,axion-review} and the timescales for these decays are
much larger than $\Lambda_{\rm QCD}^{-1}$, ie. long enough to generate
these fields but short enough to avoid cosmological problems.  In
addition, we have found some additional structures which may help
solve this problem.  We shall present these elsewhere \cite{FZ:2000}.
In any case, we assume that some mechanism exists to resolve the
domain wall problem in an appropriate timescale.  Thus, all the
relevant timescales are of the order $\Lambda_{\rm QCD}^{-1}$ except
for the lifetime of the walls which is substantially longer and thus,
although the discussed interactions will affect the qualitative
results, they will not affect the mechanism or substantially change
the order of the effects.

\section{Conclusion.}
We have shown that this mechanism can generate the magnetic fields
(\ref{*}) with large correlations.  It seems that galactic dynamos
should still play an important amplification role.  It seems that the
crucial conditions for the dynamo to take place are fields
$B>10^{-20}$ G with large ($100$ kpc) correlations.  From (\ref{*}) we
see that we have a huge interval $10^{-10}\ll\xi^*\Lambda_{\rm QCD}\leq 1$ of
$\xi^*$ to seed these dynamos.  Also, if $\xi^{*}$ is small, then this
mechanism may generate measurable extra-galactic fields.

We mention two new points that distinguish this mechanism from
previous proposals \cite{Iwazaki}.  First, the key nucleon is the
neutron which generates the fields due to an anomalous dipole moment
induced by the CP violating domain walls.  The nucleons thus make the
wall ferromagnetic, not diamagnetic as discussed in \cite{Voloshin2}.
Second, the interaction between the domain walls and nucleons are
substantial because of the similar scale ($\Lambda_{\rm QCD}^{-1}$) of
the $\eta'$ transition in the $a_{\eta'}$ domain wall.  There is no
way that axion domain walls with scales $\sim m_{a}^{-1}$ can
efficiently align nucleons at a temperature $T_{\rm QCD}$.

The presence of the magnetic fields generated by our mechanism may
have several observable effects.  First, large magnetic fields may
alter nucleosynthesis production ratios \cite{SM:1998}.  Secondly,
large scale magnetic fields may distort the CMB spectrum in a
measurable manner \cite{JKO:1999}.  These place upper bounds on the
strength of the fields.  Even the maximal fields (\ref{eq:EB}) with
$\xi^*\Lambda_{\rm QCD}=1$ generated by domain walls lie within these
bounds. Also, if $\xi^*$ turns out to be quite small, then, unless the
distribution of galaxies is correlated with the domain walls, this
mechanism might generate measurable extra-galactic fields.

Two other effects may be closely related to magnetic fields generated
from domain walls.  One is the observation of ultra-high energy cosmic
rays past the GZK cutoff \cite{GZK}.  Magnetic fields on the
scale of those discussed here may hold a key to explaining this
mystery.  The other is an apparent anisotropy of radiation propagation
over large distances resulting in a constant offset in Faraday
measurements \cite{polarization}.  One possible explanation involves
the introduction of a Chern-Simons term by hand \cite{CFJ:1990}.  This
type of term might arise naturally from CP violating domain walls.

Domain walls at the QCD phase transition provide a nice method of
generating magnetic fields on $100$ kpc correlations today
(\ref{*}).  In addition, the fields and domain walls key to this
mechanism may play a role in a number of unexplained
astrophysical phenomena.  We conclude on this optimistic note.
 
This work was supported by the NSERC of Canada.  We would like to
thank R. Brandenberger for many useful discussions.  AZ wishes to
thank: M. Shaposhnikov and I. Tkachev for valuable discussions which
motivated this study; Larry McLerran and D. Son for discussions on
Silk damping; and M. Voloshin and A. Vainshtein for discussions on the
magnetic properties of domain walls.
\section*{References}
\bibliographystyle{prsty}

\end{document}